\documentclass{article}
\usepackage{amssymb}
\usepackage{amsmath}
\usepackage{hyperref}

\input{tcilatex}
\begin{document}

\title{Massive Dual Gravity in $N$ Spacetime Dimensions}
\author{H. Alshal$^{\S \bigtriangleup \ast }$\ and T.L. Curtright$^{\S \dag
} $ \\
$^{\S }$Department of Physics, University of Miami, Coral Gables, Florida
33124\\
$^{\bigtriangleup }$Department of Physics, Faculty of Science, Cairo
University, Giza, 12613, Egypt\\
$^{\ast }${\footnotesize halshal@sci.cu.edu.eg\ \ \ } $\ ^{\dag }$%
{\footnotesize curtright@miami.edu}}
\date{}
\maketitle

\begin{abstract}
We describe a field theory for \textquotedblleft massive dual
gravity\textquotedblright\ in $N$ spacetime dimensions. \ We obtain a
Lagrangian that gives the lowest order coupling of the field to the $N$%
-dimensional curl of its own energy-momentum tensor. \ We then briefly
discuss classical solutions. \ Finally, we show the theory is the exact dual
of the Ogievetsky-Polubarinov model generalized to any $N$.

\vfill

\noindent \hrulefill
\end{abstract}

\vfill

The suggestion that the gravitational field might be massive was made long
ago and has been studied in great detail by various authors, albeit for a
very small mass with an extremely slow exponential fall-off for the
corresponding Yukawa potential (e.g. see \cite{FMS1968}). \ The subject was
surveyed almost exhaustively in \cite{P}. \ However, that survey completely
overlooked at least one interesting possibility.

Consider a field theory dual to that of a massive symmetric tensor ($h_{\mu
\nu }=h_{\nu \mu }$) in $N$ spacetime dimensions (i.e. \textquotedblleft
ND\textquotedblright ), as an extension of the ideas and results in \cite%
{C1980,C2019}. \ For ND the dual field of $h_{\mu \nu }$ is another tensor
field whose rank depends on whether the original $h_{\mu \nu }$ tensor is
massless or massive. \ If massless the dual of $h_{\mu \nu }$ is a tensor $%
T_{\left[ \lambda _{1}\cdots \lambda _{N-3}\right] \mu }$ of rank $N-2$,
while if massive the dual of $h_{\mu \nu }$ is a tensor $T_{\left[ \lambda
_{1}\cdots \lambda _{N-2}\right] \mu }$ of rank $N-1$. \ The indices for
these $T$ fields are symmetrized, in an obvious way, according to the
corresponding \href{https://en.wikipedia.org/wiki/Young_tableau}{Young
tableaux}. \ Various individual fields of this type appear in string
theories \cite{C1986}, and in \textquotedblleft M-theory\textquotedblright\
and \textquotedblleft E-theory\textquotedblright\ \cite{H,W,DHN}. \ For a
recent review of duality for gravity and higher-spin fields, with an
emphasis on massless models in higher dimensions, see \cite{D}.

As a preliminary check, the number of on-shell degrees of freedom
(\textquotedblleft $\func{dof}$\textquotedblright ) for these different
fields are as follows, when considered to be irreducible (hence traceless)
tensor representations of $O\left( N-2\right) $ and $O\left( N-1\right) $
for the massless and massive cases, respectively, as computed using the
well-known \href{https://en.wikipedia.org/wiki/Hook_length_formula}{%
hook-length} rules and the \href{https://en.wikipedia.org/wiki/Schur-3Weyl_duality}%
{Schur-Weyl theorem}.%
\begin{eqnarray}
\func{dof}\left( h_{\mu \nu }\right)  &=&\tfrac{\left( N-2\right) \left(
N-1\right) }{2}-1=\tfrac{N\left( N-3\right) }{2}\text{ \ \ if massless.} \\
\func{dof}\left( T_{\left[ \lambda _{1}\cdots \lambda _{N-3}\right] \mu
}\right)  &=&\tfrac{\left( 2\right) \cdots \left( N-2\right) \left(
N-1\right) }{\left( 1\right) \cdots \left( N-4\right) \left( N-2\right)
\left( 1\right) }-\tfrac{\left( 3\right) \cdots \left( N-2\right) }{\left(
1\right) \cdots \left( N-4\right) }=\tfrac{N\left( N-3\right) }{2}\text{ \ \
if massless.}  \notag
\end{eqnarray}%
\begin{eqnarray}
\func{dof}\left( h_{\mu \nu }\right)  &=&\tfrac{\left( N-1\right) \left(
N\right) }{2}-1=\tfrac{\left( N+1\right) \left( N-2\right) }{2}\text{ \ \ if
massive.} \\
\func{dof}\left( T_{\left[ \lambda _{1}\cdots \lambda _{N-2}\right] \mu
}\right)  &=&\tfrac{\left( 2\right) \cdots \left( N-1\right) \left( N\right) 
}{\left( 1\right) \cdots \left( N-3\right) \left( N-1\right) \left( 1\right) 
}-\tfrac{\left( 3\right) \cdots \left( N-1\right) }{\left( 1\right) \cdots
\left( N-3\right) }=\tfrac{\left( N+1\right) \left( N-2\right) }{2}\text{ \
\ if massive.}  \notag
\end{eqnarray}%
Thus the degrees of freedom agree for the respective cases. \ For the
massive situation in $N$ dimensions, the ranks and $\func{dof}$s of fields
are well-known to coincide with those for the massless situation in $N+1$
dimensions. \ On the other hand, the dynamics of the fields require less
trivial considerations.

\vfill\newpage 

For massive gravity the dual free field Lagrangian density is given by \cite%
{C1980,C2019,GKMU} 
\begin{equation}
\mathcal{L}=\mathcal{K}_{\mu }^{\ \nu }\mathcal{K}_{\nu }^{\ \mu }+\frac{%
\left( -1\right) ^{N}m^{2}}{\left( N-2\right) !}\left( T_{\left[ \lambda
_{1}\cdots \lambda _{N-2}\right] \mu }T^{\left[ \lambda _{1}\cdots \lambda
_{N-2}\right] \mu }-\left( N-2\right) T_{\left[ \lambda _{1}\cdots \lambda
_{N-3}\right] }T^{\left[ \lambda _{1}\cdots \lambda _{N-3}\right] }\right) \
,  \label{LinN}
\end{equation}%
with a choice of the overall normalization, and with the definitions%
\begin{gather}
F_{\left[ \lambda _{1}\cdots \lambda _{N-1}\right] \mu }\equiv \partial
_{\lambda _{1}}T_{\left[ \lambda _{2}\cdots \lambda _{N-1}\right] \mu }\pm
\left\{ N-2\text{ signed permutations of }\lambda \text{'s}\right\} \ , \\
K_{\mu }^{\ \nu }\equiv F_{\left[ \lambda _{1}\cdots \lambda _{N-1}\right]
\mu }\varepsilon ^{\lambda _{1}\cdots \lambda _{N-1}\nu }=\left( N-1\right)
\partial _{\lambda _{1}}T_{\left[ \lambda _{2}\cdots \lambda _{N-1}\right]
\mu }\varepsilon ^{\lambda _{1}\cdots \lambda _{N-1}\nu }\ , \\
\mathcal{K}_{\mu }^{\ \nu }\equiv \frac{1}{\left( N-1\right) !}~K_{\mu }^{\
\nu }\ ,\ \ \ T_{\left[ \lambda _{1}\cdots \lambda _{N-3}\right] }\equiv T_{%
\left[ \lambda _{1}\cdots \lambda _{N-3}\mu \right] \nu }\eta ^{\mu \nu }\ ,
\end{gather}%
where the Lorentz metric is $\eta _{\mu \nu }=\func{diag}\left( +1,-1,\cdots
,-1\right) $. \ Some $N$-dependent coefficients have been incorporated into
the definition of $\mathcal{K}_{\mu }^{\ \nu }$ to take into account the
number of \emph{antisymmetrized} summed indices in the definition of $K_{\mu
}^{\ \nu }$.

It is instructive to compare $\mathcal{L}$ to the previously studied 4D case 
\cite{C1980,C2019}. \ 
\begin{eqnarray}
\mathcal{L}_{4D} &=&-\frac{1}{6}\left( F_{\left[ \lambda \mu \nu \right]
\rho }~F^{\left[ \lambda \mu \nu \right] \rho }-3F_{\left[ \mu \nu \right]
}~F^{\left[ \mu \nu \right] }\right) +\frac{1}{2}~m^{2}\left( T_{\left[
\lambda \mu \right] \nu }T^{\left[ \lambda \mu \right] \nu }-2T_{\lambda
}T^{\lambda }\right)  \notag \\
&=&\mathcal{K}_{\mu }^{\ \nu }\mathcal{K}_{\nu }^{\ \mu }+\frac{1}{2}%
~m^{2}\left( T_{\left[ \lambda \mu \right] \nu }T^{\left[ \lambda \mu \right]
\nu }-2T_{\lambda }T^{\lambda }\right) \ .  \label{4DMassiveLagrangian}
\end{eqnarray}%
This agrees with (\ref{LinN}) for $N=4$. \ The reader should consider $N=3$
for a simpler example.

The free field equations are summarized in Appendix A. \ A consistent
interacting field equation for the massive ND model is an obvious
generalization of the 4D equation \cite{C1980,C2019}, namely,%
\begin{equation}
\left( \square +m^{2}\right) T_{\left[ \lambda _{1}\cdots \lambda _{N-2}%
\right] \nu }=\kappa P_{\lambda _{1}\cdots \lambda _{N-2}\nu ,\alpha \beta
\gamma }\partial ^{\alpha }\Theta ^{\beta \gamma }\ ,  \label{FE}
\end{equation}%
where a symmetrizer is defined to be%
\begin{eqnarray}
P_{\lambda _{1}\cdots \lambda _{N-2}\nu ,\alpha \beta \gamma } &=&\left(
N-2\right) \varepsilon _{\lambda _{1}\cdots \lambda _{N-2}\alpha \beta }\eta
_{\gamma \nu } \\
&&+\varepsilon _{\nu \lambda _{2}\cdots \lambda _{N-2}\alpha \beta }\eta
_{\gamma \lambda _{1}}+\varepsilon _{\lambda _{1}\nu \lambda _{3}\cdots
\lambda _{N-2}\alpha \beta }\eta _{\gamma \lambda _{2}}+\cdots +\varepsilon
_{\lambda _{1}\cdots \lambda _{N-3}\nu \alpha \beta }\eta _{\gamma \lambda
_{N-2}}\ ,  \notag
\end{eqnarray}%
and where $\Theta _{\mu \nu }$ is any conserved, symmetric tensor, e.g. the
energy-momentum tensor, and $\kappa $ is a dimensionful parameter with units 
$1/m^{N/2}$ since dimensionally $\left[ T\right] =\frac{1}{2}\left(
N-2\right) $ in mass units. \ It is natural to express $\kappa $ in terms of
Newton's constant in ND, and a length-scale set by the size of the
envisioned ND universe, similar to the expression in 4D \cite{C1980,C2019}.
\ The RHS of (\ref{FE}) is obtained below, to $O\left( \kappa \right) $,
from a Lagrangian. \ 

The field equation (\ref{FE}) implies that the trace $T_{\left[ \lambda
_{1}\cdots \lambda _{N-3}\right] }=\eta ^{\nu \lambda _{N-2}}T_{\left[
\lambda _{1}\cdots \lambda _{N-2}\right] \nu }$ and all divergences of $T_{%
\left[ \lambda _{1}\cdots \lambda _{N-2}\right] \nu }$ decouple, i.e. they\
are free fields, and therefore they may be consistently set to zero leaving
on-shell states that comprise only a single $SO\left( N-1\right) $
representation of mass $m$. \ For example, when $N=4$ unadulterated massive
spin 2 states are obtained on-shell.

The on-shell field equation for the $T$-field strength $\mathcal{K}_{\mu
}^{\ \nu }$ is%
\begin{equation}
\left( \square +m^{2}\right) \mathcal{K}_{\mu }^{\ \nu }=\kappa \left(
\left( 1-N\right) \square \Theta _{\mu }^{\ \nu }+\left( \delta _{\mu }^{\
\nu }\square -\partial _{\mu }\partial ^{\nu }\right) \Theta \right) \ .
\label{KE}
\end{equation}%
This on-shell result for the $\mathcal{K}$-tensor follows from (\ref{FE})
and the identity%
\begin{equation}
\varepsilon ^{\lambda \lambda _{1}\cdots \lambda _{N-2}\nu }P_{\lambda
_{1}\cdots \lambda _{N-2}\mu ,\alpha \beta \gamma }=\left( N-2\right)
!\left( \left( 2-N\right) \delta _{\alpha \beta }^{\lambda \nu }~\eta
_{\gamma \mu }+\delta _{\mu \alpha \beta }^{\lambda \sigma \nu }~\eta
_{\gamma \sigma }\right) \ .
\end{equation}%
In principle, there appear to be no fundamental barriers to prevent
obtaining the field equations (\ref{FE}) and (\ref{KE}) from a closed-form
Lagrangian for self-coupled dual fields, with the sources given to all
orders in $\kappa $. \ Such is the case for the massive dual scalar field 
\cite{C2019} (also see Appendix C). \ But it will suffice here to do this
only to lowest order in $\kappa $.

The massive free field energy-momentum tensor,%
\begin{eqnarray}
\mathcal{\theta }_{\mu }^{\ \nu } &=&\mathcal{K}_{\mu }^{\ \lambda }\mathcal{%
K}_{\lambda }^{\ \nu }+\frac{\left( -1\right) ^{N-1}m^{2}}{\left( N-3\right)
!}~T_{\left[ \mu \alpha _{2}\cdots \alpha _{N-2}\right] \lambda }T^{\left[
\nu \alpha _{2}\cdots \alpha _{N-2}\right] \lambda }  \notag \\
&&-\frac{1}{2}\ \delta _{\mu }^{\ \nu }\left( \mathcal{K}_{\alpha \beta }%
\mathcal{K}^{\beta \alpha }-\frac{\left( -1\right) ^{N}m^{2}}{\left(
N-2\right) !}~T_{\left[ \alpha _{1}\cdots \alpha _{N-2}\right] \gamma }T^{%
\left[ \alpha _{1}\cdots \alpha _{N-2}\right] \gamma }\right) \ ,
\end{eqnarray}%
is\ symmetric and conserved on-shell given the $O\left( \kappa ^{0}\right) $
field equations, as discussed in the Appendices. \ To obtain the field
equations (\ref{FE}) and (\ref{KE}) to $O\left( \kappa \right) $ this
energy-momentum tensor must be augmented by adding a manifestly\ conserved ($%
\partial ^{\mu }\vartheta _{\mu }^{\ \nu }\equiv 0$), symmetric ($\vartheta
_{\mu \nu }=\vartheta _{\nu \mu }$) \textquotedblleft
improvement\textquotedblright , namely, 
\begin{eqnarray}
\Theta _{\mu }^{\ \nu } &=&\theta _{\mu }^{\ \nu }+\frac{\left( -1\right)
^{N-1}}{\left( N-3\right) !}~\vartheta _{\mu }^{\ \nu }\ , \\
\vartheta _{\mu }^{\ \nu } &\equiv &\square \left( T_{\left[ \mu \alpha
_{2}\cdots \alpha _{N-2}\right] c}T^{\left[ \nu \alpha _{2}\cdots \alpha
_{N-2}\right] c}\right) +\delta _{\mu }^{\ \nu }\partial _{a}\partial
^{b}\left( T^{\left[ a\alpha _{2}\cdots \alpha _{N-2}\right] c}T_{\left[
b\alpha _{2}\cdots \alpha _{N-2}\right] c}\right)  \notag \\
&&-\partial _{\mu }\partial ^{b}\left( T_{\left[ b\alpha _{2}\cdots \alpha
_{N-2}\right] c}T^{\left[ \nu \alpha _{2}\cdots \alpha _{N-2}\right]
c}\right) -\partial ^{\nu }\partial _{b}\left( T^{\left[ b\alpha _{2}\cdots
\alpha _{N-2}\right] c}T_{\left[ \mu \alpha _{2}\cdots \alpha _{N-2}\right]
c}\right) \ .
\end{eqnarray}

A Lagrangian which gives the sought-for field equation to $O\left( \kappa
\right) $ (but unfortunately, \emph{not} to $O\left( \kappa ^{2}\right) $)
is then obtained by adding to the massive free field Lagrangian (\ref{LinN}) 
$O\left( \kappa \right) $ interactions \emph{suggested} by the form $%
\mathcal{K}_{\alpha }^{\ \beta }\Theta _{\beta }^{\ \alpha }$, namely,%
\begin{eqnarray}
\mathcal{L}_{int} &=&\frac{1}{3}\left( -1\right) ^{N-1}\left( N-1\right)
!~\kappa ~\mathcal{K}_{\alpha }^{\ \beta }\mathcal{K}_{\beta }^{\ \gamma }%
\mathcal{K}_{\gamma }^{\ \alpha } \\
&&+\frac{\left( -1\right) ^{N-1}\kappa }{\left( N-3\right) !}~T_{\left[
\lambda _{1}\cdots \lambda _{N-2}\right] \nu }P_{\ \ \ \ \ \ \ \ \ \ \ \ \ \
\ \ \gamma }^{\lambda _{1}\cdots \lambda _{N-2}\nu ,\alpha \beta }~\partial
_{\alpha }\left( 
\begin{array}{c}
\left( \square +m^{2}\right) \left( T_{\left[ \beta \alpha _{2}\cdots \alpha
_{N-2}\right] c}T^{\left[ \gamma \alpha _{2}\cdots \alpha _{N-2}\right]
c}\right) \\ 
\\ 
-\partial ^{\gamma }\partial _{b}\left( T^{\left[ b\alpha _{2}\cdots \alpha
_{N-2}\right] c}T_{\left[ \beta \alpha _{2}\cdots \alpha _{N-2}\right]
c}\right)%
\end{array}%
\right) \ ,  \notag
\end{eqnarray}%
up to a relative normalization between $\mathcal{L}$\ and $\mathcal{L}_{int}$%
. \ The resulting action due to $\mathcal{L}_{int}$ is of course%
\begin{equation}
\mathcal{A}_{int}=\int \mathcal{L}_{int}d^{N}x\ ,
\end{equation}%
and therefore, by varying $T^{\left[ \lambda _{1}\cdots \lambda _{N-2}\right]
\nu }$ in\ $\mathcal{A}_{int}$, the contributions to the field equations
follow from%
\begin{gather}
\delta \mathcal{A}_{int}=\frac{\left( -1\right) ^{N-1}\kappa }{\left( \left(
N-1\right) !\right) ^{2}}~\int \left( \delta K_{\alpha }^{\ \beta }\right)
K_{\beta }^{\ \gamma }K_{\gamma }^{\ \alpha }d^{N}x \\
+\frac{\left( -1\right) ^{N-1}\kappa }{\left( N-3\right) !}~\int \left(
\delta T_{\left[ \lambda _{1}\cdots \lambda _{N-2}\right] \nu }\right) P_{\
\ \ \ \ \ \ \ \ \ \ \ \ \ \ \ \gamma }^{\lambda _{1}\cdots \lambda _{N-2}\nu
,\alpha \beta }~\partial _{\alpha }\left( 
\begin{array}{c}
\left( \square +m^{2}\right) \left( T_{\left[ \beta \alpha _{2}\cdots \alpha
_{N-2}\right] c}T^{\left[ \gamma \alpha _{2}\cdots \alpha _{N-2}\right]
c}\right) \\ 
\\ 
-\partial ^{\gamma }\partial _{b}\left( T^{\left[ b\alpha _{2}\cdots \alpha
_{N-2}\right] c}T_{\left[ \beta \alpha _{2}\cdots \alpha _{N-2}\right]
c}\right)%
\end{array}%
\right) d^{N}x  \notag \\
+\frac{\left( -1\right) ^{N-1}\kappa }{\left( N-3\right) !}~\int T_{\left[
\lambda _{1}\cdots \lambda _{N-2}\right] \nu }P_{\ \ \ \ \ \ \ \ \ \ \ \ \ \
\ \ \gamma }^{\lambda _{1}\cdots \lambda _{N-2}\nu ,\alpha \beta }~\partial
_{\alpha }\left( 
\begin{array}{c}
\left( \square +m^{2}\right) \delta \left( T_{\left[ \beta \alpha _{2}\cdots
\alpha _{N-2}\right] c}T^{\left[ \gamma \alpha _{2}\cdots \alpha _{N-2}%
\right] c}\right) \\ 
\\ 
-\partial ^{\gamma }\partial _{b}\delta \left( T^{\left[ b\alpha _{2}\cdots
\alpha _{N-2}\right] c}T_{\left[ \beta \alpha _{2}\cdots \alpha _{N-2}\right]
c}\right)%
\end{array}%
\right) d^{N}x  \notag
\end{gather}%
Upon integrating by parts the terms in the last line give \emph{no}
contributions to the bulk field equations at $O\left( \kappa \right) $
because of the $O\left( \kappa ^{0}\right) $ on-shell conditions (cf. (\ref%
{OnShell3}) and (\ref{KG}) in Appendix A). \ These terms are important at $%
O\left( \kappa ^{2}\right) $, but they have no effect at $O\left( \kappa
\right) $.

After integrating by parts, the bulk variation of the $K$ trilinear becomes 
\begin{gather}
\frac{\left( -1\right) ^{N-1}\kappa }{\left( \left( N-1\right) !\right) ^{2}}%
\int \left( \delta K_{\alpha }^{\ \beta }\right) K_{\beta }^{\ \gamma
}K_{\gamma }^{\ \alpha }d^{4}x=\frac{\left( -1\right) ^{N-2}\left(
N-1\right) \kappa }{\left( \left( N-1\right) !\right) ^{2}}\int \left(
\delta T_{\left[ \alpha _{2}\cdots \alpha _{N-1}\right] \alpha }\right)
\varepsilon ^{\alpha _{1}\cdots \alpha _{N-1}\beta }\partial _{\alpha
_{1}}\left( K_{\beta }^{\ \gamma }K_{\gamma }^{\ \alpha }\right) d^{N}x 
\notag \\
=\frac{\kappa }{\left( \left( N-1\right) !\right) ^{2}}~\int \left( \delta
T_{\left[ \lambda _{1}\cdots \lambda _{N-2}\right] \nu }\right) P_{\ \ \ \ \
\ \ \ \ \ \ \ \ \ \ \ \gamma }^{\lambda _{1}\cdots \lambda _{N-2}\nu ,\alpha
\beta }~\partial _{\alpha }\left( K_{\beta }^{\ \mu }K_{\mu }^{\ \gamma
}\right) d^{N}x
\end{gather}%
where we have also exploited the symmetry of $T_{\left[ \alpha _{2}\cdots
\alpha _{N-1}\right] \alpha }$ and that of the symmetrizer to write%
\begin{equation}
\left( N-1\right) \left( \delta T_{\left[ \alpha _{2}\cdots \alpha _{N-1}%
\right] \alpha }\right) \varepsilon ^{\alpha _{1}\cdots \alpha _{N-1}\beta
}\eta ^{\alpha \gamma }\partial _{\alpha _{1}}=\left( -1\right) ^{N-2}\left(
\delta T_{\left[ \lambda _{1}\cdots \lambda _{N-2}\right] \nu }\right)
P^{\lambda _{1}\cdots \lambda _{N-2}\nu ,\alpha \beta \gamma }~\partial
_{\alpha }
\end{equation}%
The $O\left( \kappa \right) $ variation of the interaction is therefore%
\begin{gather}
\delta \mathcal{A}_{int}=\kappa \int \left( \delta T_{\left[ \lambda
_{1}\cdots \lambda _{N-2}\right] \nu }\right) P_{\ \ \ \ \ \ \ \ \ \ \ \ \ \
\ \ \gamma }^{\lambda _{1}\cdots \lambda _{N-2}\nu ,\alpha \beta }~\partial
_{\alpha }\left( \mathcal{K}_{\beta }^{\ \mu }\mathcal{K}_{\mu }^{\ \gamma
}\right) d^{N}x \\
+\frac{\left( -1\right) ^{N-1}\kappa }{\left( N-3\right) !}~\int \left(
\delta T_{\left[ \lambda _{1}\cdots \lambda _{N-2}\right] \nu }\right) P_{\
\ \ \ \ \ \ \ \ \ \ \ \ \ \ \ \gamma }^{\lambda _{1}\cdots \lambda _{N-2}\nu
,\alpha \beta }~\partial _{\alpha }\left( 
\begin{array}{c}
\left( \square +m^{2}\right) \left( T_{\left[ \beta \alpha _{2}\cdots \alpha
_{N-2}\right] c}T^{\left[ \gamma \alpha _{2}\cdots \alpha _{N-2}\right]
c}\right) \\ 
\\ 
-\partial ^{\gamma }\partial _{b}\left( T^{\left[ b\alpha _{2}\cdots \alpha
_{N-2}\right] c}T_{\left[ \beta \alpha _{2}\cdots \alpha _{N-2}\right]
c}\right)%
\end{array}%
\right) d^{N}x  \notag \\
+\frac{\left( -1\right) ^{N-1}\kappa }{\left( N-3\right) !}~\int T_{\left[
\lambda _{1}\cdots \lambda _{N-2}\right] \nu }~P_{\ \ \ \ \ \ \ \ \ \ \ \ \
\ \ \ \gamma }^{\lambda _{1}\cdots \lambda _{N-2}\nu ,\alpha \beta
}~\partial _{\alpha }\left( 
\begin{array}{c}
\left( \square +m^{2}\right) \delta \left( T_{\left[ \beta \alpha _{2}\cdots
\alpha _{N-2}\right] c}T^{\left[ \gamma \alpha _{2}\cdots \alpha _{N-2}%
\right] c}\right) \\ 
\\ 
-\partial ^{\gamma }\partial _{b}\delta \left( T^{\left[ b\alpha _{2}\cdots
\alpha _{N-2}\right] c}T_{\left[ \beta \alpha _{2}\cdots \alpha _{N-2}\right]
c}\right)%
\end{array}%
\right) d^{N}x  \notag
\end{gather}%
That is to say,%
\begin{equation}
\delta \mathcal{A}_{int}=\kappa \int \left( \delta T^{\left[ \lambda
_{1}\cdots \lambda _{N-2}\right] \nu }\right) P_{\lambda _{1}\cdots \lambda
_{N-2}\nu ,\alpha \beta \gamma }\partial ^{\alpha }\Theta ^{\beta \gamma
}d^{N}x+O\left( \kappa ^{2}\right)
\end{equation}%
This variation thereby gives precisely the RHS of the field equation (\ref%
{FE}) to lowest non-trivial order in $\kappa $.

Given that the RHS of (\ref{FE}) is a total divergence, it may be somewhat
surprising that energy-momentum \emph{can} produce dual fields that are 
\emph{in}distinguishable from conventional massive gravity solutions
\textquotedblleft outside the source\textquotedblright\ especially in the
weak-field limit where the energy-momentum is due to sources other than the $%
T$-field itself. \ \ This is perhaps more easily seen from (\ref{KE}). \ In
fact, that field equation is closely related to other, more familiar
expressions. \ 

Were it not for the manifestly conserved trace term, $\left( \partial _{\mu
}\partial _{\nu }-\eta _{\mu \nu }\square \right) \Theta $, an obvious but
naive inference from (\ref{KE}) would be that a more conventional form of
massive gravity, such as that in \cite{FMS1968}, would be related to the
on-shell dual theory just by the local identification\footnote{%
A local identification $K_{\mu \nu }\left( x\right) \propto \square h_{\mu
\nu }\left( x\right) $ would require a less palatable \textit{non}local
inverse relation, $h_{\mu \nu }\left( x\right) \propto h_{\mu \nu }^{\left(
0\right) }\left( x\right) +\int G\left( x,y\right) K_{\mu \nu }\left(
y\right) d^{N}y$ where $G$ is a Green function such that $\square G\left(
x,y\right) =\delta ^{N}\left( x-y\right) $ and $h_{\mu \nu }^{\left(
0\right) }$ is a free massless field.} $K_{\mu \nu }\propto \square h_{\mu
\nu }$, where%
\begin{equation}
\left( \square +m^{2}\right) h_{\mu \nu }=\kappa \Theta _{\mu \nu }\ .
\label{CFE}
\end{equation}%
The trace term invalidates this simple identification, in general. \
Nevertheless, there \emph{are} situations where the dual and conventional
theories give equivalent results. \ This is especially true for static
configurations.

Static sources do indeed produce $\mathcal{K}_{00}$ fields. \ In the
weak-field limit where $T$-field dependence in $\Theta _{\mu \nu }$ can be
ignored, the static equation is%
\begin{equation}
\left( \nabla ^{2}-m^{2}\right) \mathcal{K}_{00}=-\kappa \nabla ^{2}\left(
\left( N-1\right) \Theta _{00}-\Theta \right) \ ,
\end{equation}%
an inhomogeneous equation with well-known solutions, for given static
sources on the RHS. \ That is to say, 
\begin{equation}
\left( \nabla ^{2}-m^{2}\right) \mathcal{K}=-\mathcal{F}\ ,
\end{equation}%
where $\mathcal{K}$ and $\mathcal{F}$\ are defined by%
\begin{equation}
\mathcal{K=}\left( \mathcal{K}_{00}+\mathcal{F}\right) /m^{2}\ ,\ \ \ 
\mathcal{F}=\kappa \left( N-1\right) \Theta _{00}-\kappa \Theta \ .
\end{equation}%
In regions where $\mathcal{F}=0$ (i.e. outside the source) then $\mathcal{K}%
\propto \mathcal{K}_{00}$. \ 

Therefore, modulo boundary conditions, the solution for $\mathcal{K}$ would
be the same as that for more conventional massive gravity, for an equivalent
conventional source, namely, for $\kappa \Theta _{00}=\mathcal{F}$. \ Thus,
outside the source in regions where $\mathcal{F}=0$, $h_{00}$ and $\mathcal{K%
}_{00}$\ could easily be indistinguishable in the weak-field limit.

If $\Theta =0$ then clearly this indistinguishability could carry over to
more general situations, including those with time dependence, since for
vanishing energy-momentum trace,%
\begin{equation}
\left( \square +m^{2}\right) \mathcal{H}_{\mu \nu }=\kappa \Theta _{\mu \nu
}\ ,  \label{HE}
\end{equation}%
with the field redefinition%
\begin{equation}
\mathcal{H}_{\mu \nu }=\frac{1}{m^{2}\left( N-1\right) }\left( \mathcal{K}%
_{\mu \nu }+\kappa \left( N-1\right) \Theta _{\mu \nu }\right) \ .
\label{HDefn}
\end{equation}%
So if $\Theta =0$ the field equation for $\mathcal{H}_{\mu \nu }$ coincides
with that for $h_{\mu \nu }$. \ In this case, with suitable boundary
conditions, the solutions would again be the same.

The preceding remarks suggest that (\ref{HDefn}) may be useful for the dual
theory even when $\Theta \neq 0$ and even when the energy-momentum tensor
includes contributions from the $T$-field itself so that the weak-field
limit does not apply. \ In that case (\ref{HDefn}) is a \emph{nonlinear} 
\emph{field redefinition} that leads to the following equivalent restatement
of (\ref{KE}). 
\begin{equation}
\left( \square +m^{2}\right) \mathcal{H}_{\mu \nu }=\kappa \Theta _{\mu \nu
}+\frac{\kappa }{\left( N-1\right) m^{2}}\left( \eta _{\mu \nu }\square
-\partial _{\mu }\partial _{\nu }\right) \Theta \ .  \label{OPEND}
\end{equation}%
From the first kinematic constraint in (\ref{Kinematic}), the trace $%
\mathcal{H}=\mathcal{H}_{\mu }^{\ \mu }$ is then fixed by (\ref{HDefn}) to be%
\begin{equation}
\mathcal{H}=\frac{\kappa }{m^{2}}~\Theta \ .  \label{TraceH}
\end{equation}%
This constraint on the trace is consistent with (\ref{OPEND}) because, given
that field equation, the difference $\mathcal{H}-\kappa \Theta /m^{2}$ is a
free field. \ Similarly, for conserved and symmetric $\Theta _{\mu \nu }$
both the divergence and antisymmetric parts of the $\mathcal{H}$-field are
free and consistently set to zero.

As stated above, (\ref{HDefn}) is in general a nonlinear field redefinition,
given that $\Theta _{\mu \nu }$ will in general depend on the dual field,
but in the weak-field limit, outside any non-$T$-field source of
energy-momentum, the $\mathcal{H}_{\mu \nu }$ field is just proportional to $%
\mathcal{K}_{\mu \nu }$, hence proportional to the $T$-field strength. \
This is an expected relation that characterizes massive free (or weak) field
duality: \ Field and field strength are interchanged \cite{C1980}.

More importantly, as previously noted for the 4D case \cite{C2019}, the
field equation (\ref{OPEND}) is \emph{not} the conventional one in (\ref{CFE}%
). \ That is to say, (\ref{FE}) and (\ref{CFE}) are \emph{not} massive duals
of one another, in general. \ Rather, (\ref{OPEND}) is the ND extension of
the field equation proposed by Ogievetsky and Polubarinov for a purely spin
2 massive field in 4D \cite{OP}. \ In that model $\mathcal{H}_{\mu \nu }$
would play the role of an elementary field, whereas in the theory described
here $\mathcal{H}_{\mu \nu }$ is essentially the field strength of the dual $%
T$-field, albeit with some nonlinear embellishments due to the interaction.
\ That is to say, the ND interacting massive $T$-theory described here is
the exact dual of the ND Ogievetsky-Polubarinov model, with on-shell
equivalence specified by (\ref{HDefn}).

More complete discussion of the phenomenological differences between the
dual model given here and other massive gravity fields, for realistic source
terms and sufficiently small values of $m^{2}$, will be given
elsewhere.\bigskip

\noindent \textbf{Acknowledgements} \ We thank T.S. Van Kortryk for
discussions, and especially for his concise contribution to Appendix C. \
This work was supported in part by a University of Miami Cooper Fellowship.

\section*{Appendix A: \ Dual free field equations}

The\ bulk variation of the dual free field action is%
\begin{gather}
\int \delta \mathcal{L}~d^{N}x=\int \frac{1}{\left( \left( N-1\right)
!\right) ^{2}}~\delta \left( K_{\mu \nu }K^{\nu \mu }\right) d^{N}x  \tag{A1}
\\
+\int \frac{\left( -1\right) ^{N}m^{2}}{\left( N-2\right) !}~\delta \left(
T_{\left[ \alpha _{1}\cdots \alpha _{N-2}\right] \mu }T^{\left[ \alpha
_{1}\cdots \alpha _{N-2}\right] \mu }-\left( N-2\right) T_{\left[ \alpha
_{1}\cdots \alpha _{N-3}\right] }T^{\left[ \alpha _{1}\cdots \alpha _{N-3}%
\right] }\right) d^{N}x  \notag
\end{gather}%
where we define\footnote{%
NB $\ K_{\mu }^{\ \lambda }=\left( N-1\right) !~\mathcal{K}_{\mu }^{\
\lambda }$. \ Our rationale for using $K$ as well as $\mathcal{K}$\ is to be
consistent with the notation in \cite{C2019}.}%
\begin{equation}
K_{\mu }^{\ \lambda }=F_{\left[ \lambda _{1}\cdots \lambda _{N-1}\right] \mu
}\varepsilon ^{\lambda _{1}\cdots \lambda _{N-1}\lambda }=\left( N-1\right)
\partial _{\alpha _{1}}T_{\left[ \alpha _{2}\cdots \alpha _{N-1}\right] \mu
}\varepsilon ^{\alpha _{1}\cdots \alpha _{N-1}\lambda }\ .  \tag{A2}
\label{KDefn}
\end{equation}%
That is to say,%
\begin{eqnarray}
&&\int \delta \mathcal{L}~d^{N}x  \TCItag{A3} \\
&=&\frac{2\left( -1\right) ^{N-1}}{\left( \left( N-1\right) !\right) ^{2}}%
\int \left( \delta T_{\left[ \alpha _{1}\cdots \alpha _{N-2}\right] \sigma
}\right) P^{\left[ \alpha _{1}\cdots \alpha _{N-2}\right] \sigma ,\lambda
\mu \nu }\left( \partial _{\lambda }K_{\mu \nu }+\left( N-1\right) \left(
-1\right) ^{N-1}m^{2}\varepsilon _{\nu \lambda \omega _{1}\cdots \omega
_{N-2}}T_{\ \ \ \ \ \ \ \ \ \ \ \ \ \mu }^{\left[ \omega _{1}\cdots \omega
_{N-2}\right] }\right) d^{N}x  \notag
\end{eqnarray}%
Hence the dual free field equations, in raw form, are%
\begin{equation}
P^{\left[ \alpha _{1}\cdots \alpha _{N-2}\right] \sigma ,\lambda \mu \nu
}\left( \partial _{\lambda }K_{\mu \nu }+\left( N-1\right) \left( -1\right)
^{N-1}m^{2}\varepsilon _{\nu \lambda \omega _{1}\cdots \omega _{N-2}}T_{\ \
\ \ \ \ \ \ \ \ \ \ \ \mu }^{\left[ \omega _{1}\cdots \omega _{N-2}\right]
}\right) =0\ ,  \tag{A4}  \label{RawFE}
\end{equation}%
supplemented by the kinematic conditions,%
\begin{equation}
K_{\mu }^{\ \mu }=0\ ,\ \ \ \partial _{\nu }K_{\mu }^{\ \nu }=0\ .  \tag{A5}
\label{Kinematic}
\end{equation}

By taking various divergences and contractions of (\ref{RawFE}), the field
equations boil down to the following simplified \textquotedblleft
on-shell\textquotedblright\ conditions: \ The\ Klein-Gordon equation,%
\begin{equation}
\left( \square +m^{2}\right) T_{\left[ \lambda _{1}\cdots \lambda _{N-2}%
\right] \mu }=0\ ,  \tag{A6}  \label{KG}
\end{equation}%
and the \textquotedblleft half-shell\textquotedblright\ conditions, 
\begin{eqnarray}
T_{\left[ \lambda _{1}\cdots \lambda _{N-3}\mu \right] \nu }\eta ^{\mu \nu }
&=&0\ ,  \TCItag{A7}  \label{OnShell1} \\
\partial _{\mu }T^{\left[ \mu \lambda _{2}\cdots \lambda _{N-2}\right] \nu }
&=&0\ ,  \TCItag{A8}  \label{OnShell2} \\
\partial _{\mu }T^{\left[ \lambda _{1}\cdots \lambda _{N-2}\right] \mu }
&=&0\ .  \TCItag{A9}  \label{OnShell3}
\end{eqnarray}%
Some immediate consequences of the half-shell conditions are\footnote{%
About the notation: \ As used in \cite{C2019},\ \textquotedblleft $\bumpeq $%
\textquotedblright\ means equality given one or more of the half-shell
conditions, while \textquotedblleft $\Bumpeq $\textquotedblright\ means
\textquotedblleft full-shell\textquotedblright\ equality given the
Klein-Gordon equation in addition to the half-shell conditions.}: 
\begin{equation}
\partial ^{\nu }F_{\left[ \lambda _{1}\cdots \lambda _{N-2}\mu \right] \nu
}\bumpeq 0\ ,\ \ \ F_{\left[ \lambda _{1}\cdots \lambda _{N-2}\mu \right]
\nu }\eta ^{\mu \nu }\bumpeq 0\ ,\ \ \ K_{\mu \nu }\bumpeq K_{\nu \mu }\ ,\
\ \ \partial ^{\mu }K_{\mu \nu }\bumpeq 0\ .  \tag{A10}  \label{OnShell4}
\end{equation}%
\medskip For example, the third relation in (\ref{OnShell4}) follows from
the second since 
\begin{equation}
K^{\mu \nu }-K^{\nu \mu }=\left( -1\right) ^{N-1}\left( F_{\left[ \lambda
_{1}\cdots \lambda _{N-1}\right] }^{\ \ \ \ \ \ \ \ \ \ \ \ \mu
}~\varepsilon ^{\nu \lambda _{1}\cdots \lambda _{N-1}}-F_{\left[ \lambda
_{1}\cdots \lambda _{N-1}\right] }^{\ \ \ \ \ \ \ \ \ \ \ \ \nu
}~\varepsilon ^{\mu \lambda _{1}\cdots \lambda _{N-1}}\right) \ ,  \tag{A11}
\end{equation}%
and then by ND$\ $\text{\href{http://mathworld.wolfram.com/Syzygy.html}{%
\text{syzygy}},}%
\begin{equation}
\left( -1\right) ^{N-1}\left( F_{\left[ \lambda _{1}\cdots \lambda _{N-1}%
\right] }^{\ \ \ \ \ \ \ \ \ \ \ \ \mu }~\varepsilon ^{\nu \lambda
_{1}\cdots \lambda _{N-1}}-F_{\left[ \lambda _{1}\cdots \lambda _{N-1}\right]
}^{\ \ \ \ \ \ \ \ \ \ \ \ \nu }~\varepsilon ^{\mu \lambda _{1}\cdots
\lambda _{N-1}}\right) =\left( N-1\right) F_{\left[ \lambda _{1}\cdots
\lambda _{N-2}\lambda \right] }^{\ \ \ \ \ \ \ \ \ \ \ \ \ \lambda
}~\varepsilon ^{\mu \nu \lambda _{1}\cdots \lambda _{N-2}}\text{\ .} 
\tag{A12}
\end{equation}

\section*{Appendix B: \ Free field energy-momentum conservation}

Conservation of $\mathcal{\theta }_{\mu }^{\ \nu }$ for the free theory is
established by the following Lemmata.\medskip

\noindent \textbf{[Lemma 1]}%
\begin{equation}
\partial ^{\mu }\left( K_{\mu }^{\ \lambda }K_{\lambda }^{\ \nu }-\frac{1}{2}%
\ \delta _{\mu }^{\ \nu }K_{\alpha \beta }K^{\beta \alpha }\right) \bumpeq
\left( -1\right) ^{N-2}\left( N-1\right) K_{\mu }^{\ \lambda }\varepsilon
^{\alpha _{1}\cdots \alpha _{N-2}\mu \nu }\square T_{\left[ \alpha
_{1}\cdots \alpha _{N-2}\right] \lambda }\ .  \tag{B1}  \label{L1}
\end{equation}%
Proof:%
\begin{eqnarray*}
\partial ^{\mu }\left( K_{\mu }^{\ \lambda }K_{\lambda }^{\ \nu }\right)
&=&K_{\lambda }^{\ \nu }\partial ^{\mu }K_{\mu }^{\ \lambda }+K_{\mu }^{\
\lambda }\partial ^{\mu }K_{\lambda }^{\ \nu }\bumpeq K_{\mu }^{\ \lambda
}\partial ^{\mu }K_{\lambda }^{\ \nu }=K_{\mu }^{\ \lambda }\varepsilon
^{\alpha _{1}\cdots \alpha _{N-1}\nu }\partial ^{\mu }F_{\left[ \alpha
_{1}\cdots \alpha _{N-1}\right] \lambda }\ \ \ \text{using (\ref{OnShell4})
and (\ref{KDefn})} \\
&=&K_{\mu }^{\ \lambda }\left( \varepsilon ^{\alpha _{1}\cdots \alpha
_{N-1}\mu }\partial ^{\nu }+\left( N-1\right) \varepsilon ^{\alpha
_{1}\cdots \alpha _{N-2}\mu \nu }\partial ^{\alpha _{N-1}}\right) F_{\left[
\alpha _{1}\cdots \alpha _{N-1}\right] \lambda }\ \ \ \text{\text{\href{http://mathworld.wolfram.com/Syzygy.html}%
{\text{syzygy}}} in ND \cite{Hamermashed}} \\
&\bumpeq &K_{\mu }^{\ \lambda }\partial ^{\nu }K_{\lambda }^{\ \mu }+\left(
-1\right) ^{N-2}\left( N-1\right) K_{\mu }^{\ \lambda }\varepsilon ^{\alpha
_{1}\cdots \alpha _{N-2}\mu \nu }\square T_{\left[ \alpha _{1}\cdots \alpha
_{N-2}\right] \lambda }
\end{eqnarray*}%
where in the last step we have used (\ref{KDefn}) and (\ref{OnShell2},\ref%
{OnShell3}). \ So (\ref{L1}) is established. \ Thus we are led to\medskip

\noindent \textbf{[Lemma 2]}%
\begin{equation}
K_{\mu }^{\ \lambda }\varepsilon ^{\alpha _{1}\cdots \alpha _{N-2}\mu \nu
}\bumpeq \left( -1\right) ^{N}\left( N-1\right) !~F^{\left[ \alpha
_{1}\cdots \alpha _{N-2}\nu \right] \lambda }  \tag{B2}  \label{L2}
\end{equation}%
Proof:%
\begin{eqnarray*}
K_{\mu }^{\ \lambda }\varepsilon ^{\alpha _{1}\cdots \alpha _{N-2}\mu \nu }
&\bumpeq &K_{\ \mu }^{\lambda }\varepsilon ^{\alpha _{1}\cdots \alpha
_{N-2}\mu \nu }=F^{\left[ \lambda _{1}\cdots \lambda _{N-1}\right] \lambda
}\varepsilon _{\lambda _{1}\cdots \lambda _{N-1}\mu }\varepsilon ^{\alpha
_{1}\cdots \alpha _{N-2}\mu \nu } \\
&=&\left( -1\right) ^{N}\delta _{\lambda _{1}\cdots \ \lambda
_{N-1}}^{\alpha _{1}\cdots \alpha _{N-2}\nu }F^{\left[ \lambda _{1}\cdots
\lambda _{N-1}\right] \lambda }=\left( -1\right) ^{N}\left( N-1\right) !F^{%
\left[ \alpha _{1}\cdots \alpha _{N-2}\nu \right] \lambda }
\end{eqnarray*}%
So (\ref{L2})\ is also established. \ Now, combining (\ref{L1})\ and (\ref%
{L2})\ along with (\ref{KG}) gives immediately\medskip

\noindent \textbf{[Lemma 3]}%
\begin{equation}
\partial ^{\mu }\left( K_{\mu }^{\ \lambda }K_{\lambda }^{\ \nu }-\frac{1}{2}%
\ \delta _{\mu }^{\ \nu }K_{\alpha \beta }K^{\beta \alpha }\right) \Bumpeq
-\left( N-1\right) \left( N-1\right) !~m^{2}F^{\left[ \alpha _{1}\cdots
\alpha _{N-2}\nu \right] \lambda }T_{\left[ \alpha _{1}\cdots \alpha _{N-2}%
\right] \lambda }\ .  \tag{B3}  \label{L3}
\end{equation}%
This leads to a final\medskip

\noindent \textbf{[Lemma 4]}%
\begin{equation}
F^{\left[ \alpha _{1}\cdots \alpha _{N-2}\nu \right] \lambda }T_{\left[
\alpha _{1}\cdots \alpha _{N-2}\right] \lambda }=\left( -1\right) ^{N}\left(
\partial ^{\nu }\left( \frac{1}{2}~T_{\left[ \alpha _{1}\cdots \alpha _{N-2}%
\right] \gamma }T^{\left[ \alpha _{1}\cdots \alpha _{N-2}\right] \gamma
}\right) -\left( N-2\right) \partial ^{\mu }\left( T_{\left[ \mu \alpha
_{2}\cdots \alpha _{N-2}\right] \lambda }T^{\left[ \nu \alpha _{2}\cdots
\alpha _{N-2}\right] \lambda }\right) \right)  \tag{B4}  \label{L4}
\end{equation}%
Proof:%
\begin{gather*}
F^{\left[ \alpha _{1}\cdots \alpha _{N-2}\nu \right] \lambda }T_{\left[
\alpha _{1}\cdots \alpha _{N-2}\right] \lambda }=\left( \left( -1\right)
^{N-2}\partial ^{\nu }T^{\left[ \alpha _{1}\cdots \alpha _{N-2}\right]
\lambda }+\left( N-2\right) \partial ^{\alpha _{1}}T^{\left[ \alpha
_{2}\cdots \alpha _{N-2}\nu \right] \lambda }\right) T_{\left[ \alpha
_{1}\cdots \alpha _{N-2}\right] \lambda }\ \ \ \text{definition of }F \\
\bumpeq \partial ^{\nu }\left( \frac{1}{2}\left( -1\right) ^{N}T_{\left[
\alpha _{1}\cdots \alpha _{N-2}\right] \gamma }T^{\left[ \alpha _{1}\cdots
\alpha _{N-2}\right] \gamma }\right) +\left( N-2\right) \left( -1\right)
^{N-3}\partial ^{\mu }\left( T_{\left[ \mu \alpha _{2}\cdots \alpha _{N-2}%
\right] \lambda }T^{\left[ \nu \alpha _{2}\cdots \alpha _{N-2}\right]
\lambda }\right) \ \ \ \text{using (\ref{OnShell2})}
\end{gather*}%
So (\ref{L4}) is established.

Combining (\ref{L3})\ and (\ref{L4})\ we then obtain%
\begin{eqnarray}
&&\partial ^{\mu }\left( K_{\mu }^{\ \lambda }K_{\lambda }^{\ \nu }-\frac{1}{%
2}\ \delta _{\mu }^{\ \nu }K_{\alpha \beta }K^{\beta \alpha }\right) 
\TCItag{B5} \\
&\Bumpeq &-\left( -1\right) ^{N}\left( N-1\right) \left( N-1\right)
!m^{2}\left( \partial ^{\nu }\left( \frac{1}{2}~T_{\left[ \alpha _{1}\cdots
\alpha _{N-2}\right] \gamma }T^{\left[ \alpha _{1}\cdots \alpha _{N-2}\right]
\gamma }\right) -\left( N-2\right) \partial ^{\mu }\left( T_{\left[ \mu
\alpha _{2}\cdots \alpha _{N-2}\right] \lambda }T^{\left[ \nu \alpha
_{2}\cdots \alpha _{N-2}\right] \lambda }\right) \right) \ .  \notag
\end{eqnarray}%
That is to say, $\partial ^{\mu }\mathcal{\theta }_{\mu }^{\ \nu }\Bumpeq 0$
with $\mathcal{\theta }_{\mu }^{\ \nu }$ given by%
\begin{eqnarray}
\left( \left( N-1\right) !\right) ^{2}~\mathcal{\theta }_{\mu }^{\ \nu }
&=&K_{\mu }^{\ \lambda }K_{\lambda }^{\ \nu }-\left( N-1\right) \left(
N-2\right) \left( N-1\right) !~\left( -1\right) ^{N}m^{2}\left( T_{\left[
\mu \alpha _{2}\cdots \alpha _{N-2}\right] \lambda }T^{\left[ \nu \alpha
_{2}\cdots \alpha _{N-2}\right] \lambda }\right)  \notag \\
&&-\frac{1}{2}\ \delta _{\mu }^{\ \nu }\left( K_{\alpha \beta }K^{\beta
\alpha }-\left( N-1\right) \left( N-1\right) !~\left( -1\right) ^{N}m^{2}T_{%
\left[ \alpha _{1}\cdots \alpha _{N-2}\right] \gamma }T^{\left[ \alpha
_{1}\cdots \alpha _{N-2}\right] \gamma }\right)  \TCItag{B6}
\end{eqnarray}%
The $N$-dependent factors make a little more sense when $\theta _{\mu }^{\
\nu }$ is expressed in terms of $\mathcal{K}_{\mu }^{\ \nu }$:%
\begin{eqnarray}
\mathcal{\theta }_{\mu }^{\ \nu } &=&\mathcal{K}_{\mu }^{\ \lambda }\mathcal{%
K}_{\lambda }^{\ \nu }-\frac{\left( -1\right) ^{N}m^{2}}{\left( N-3\right) !}%
~T_{\left[ \mu \alpha _{2}\cdots \alpha _{N-2}\right] \lambda }T^{\left[ \nu
\alpha _{2}\cdots \alpha _{N-2}\right] \lambda }  \notag \\
&&-\frac{1}{2}\ \delta _{\mu }^{\ \nu }\left( \mathcal{K}_{\alpha \beta }%
\mathcal{K}^{\beta \alpha }-\frac{\left( -1\right) ^{N}m^{2}}{\left(
N-2\right) !}~T_{\left[ \alpha _{1}\cdots \alpha _{N-2}\right] \gamma }T^{%
\left[ \alpha _{1}\cdots \alpha _{N-2}\right] \gamma }\right)  \TCItag{B7}
\end{eqnarray}%
But in any case, up to an overall numerical factor, all this agrees with the
4D results \cite{C1980,C2019} when $N=4$:%
\begin{equation}
\left. \mathcal{\theta }_{\mu }^{\ \nu }\right\vert _{N=4}\propto K_{\mu
\alpha }K^{\alpha \nu }-36m^{2}T_{\left[ \mu \beta \right] \gamma }T^{\left[
\nu \beta \right] \gamma }-\delta _{\mu }^{\ \nu }\left( \frac{1}{2}%
~K_{\alpha \beta }K^{\beta \alpha }-9m^{2}T_{\left[ \alpha \beta \right]
\gamma }T^{\left[ \alpha \beta \right] \gamma }\right) \ .  \tag{B8}
\end{equation}

\newpage

\section*{Appendix C: \ Coupling a dual scalar field to $\Theta $}

In this Appendix some 4D results for scalar fields \cite{C2019} are
generalized to ND. \ 

Consider a Lagrangian density $\mathcal{L}$ depending on a vector field $%
V^{\mu }$ through the two scalar variables, 
\begin{equation}
B=V_{\mu }V^{\mu }\ ,\ \ \ F=\partial _{\mu }V^{\mu }\ .  \tag{C1}
\end{equation}%
The bulk field equations that follow from the action of $\mathcal{L}$ by
varying $V^{\mu }$ are simply%
\begin{equation}
\partial _{\mu }\mathcal{L}_{F}=2V_{\mu }\mathcal{L}_{B}\ ,  \tag{C2}
\label{FieldEqns}
\end{equation}%
where the partial derivatives of $\mathcal{L}$\ are designated by $\mathcal{L%
}_{B}\equiv \partial \mathcal{L}\left( B,F\right) /\partial B$ and $\mathcal{%
L}_{F}\equiv \partial \mathcal{L}\left( B,F\right) /\partial F$. \ 

The vector field $V_{\mu }$ is to be understood as the $N$-dimensional
spacetime dual of a totally antisymmetric, rank $N-1$, tensor gauge field, $%
V_{\alpha _{1}\cdots \alpha _{N-1}}$, with its corresponding totally
antisymmetric, gauge invariant field strength, $F_{\mu \alpha _{1}\cdots
\alpha _{N-1}}=\partial _{\mu }V_{\alpha _{1}\cdots \alpha _{N-1}}\pm N-1$
terms. \ Thus 
\begin{equation}
V^{\mu }=\frac{1}{\left( N-1\right) !}~\varepsilon ^{\mu \alpha _{1}\cdots
\alpha _{N-1}}V_{\alpha _{1}\cdots \alpha _{N-1}}\ ,\ \ \ \partial _{\mu
}V^{\mu }=\frac{1}{N!}~\varepsilon ^{\alpha _{1}\cdots \alpha _{N}}F_{\alpha
_{1}\cdots \alpha _{N}}\ .  \tag{C3}  \label{VandF}
\end{equation}%
Under massive field duality \cite{C1980}, this field strength should become
the gradient of a scalar $\Phi $, 
\begin{equation}
V_{\mu }=\partial _{\mu }\Phi \ ,  \tag{C4}  \label{VectorAsGrad}
\end{equation}%
such that 
\begin{equation}
\partial _{\mu }V_{\lambda }=\partial _{\lambda }V_{\mu }\ ,  \tag{C5}
\label{GradientCond}
\end{equation}%
The goal here is to find an $\mathcal{L}$ such that field equations for $%
V_{\mu }$ amount to (\ref{GradientCond}) along with the \textquotedblleft
simple, indeed elegant\textquotedblright\ statement \cite{C1980},%
\begin{equation}
\left( \square +m^{2}\right) V_{\mu }=\kappa ~\partial _{\mu }\Theta \ , 
\tag{C6}  \label{Elegant}
\end{equation}%
where $\Theta $ is the trace of the energy-momentum tensor for the $V$-field.

For simplicity, suppose $\mathcal{L}_{B}=a+b\mathcal{L}_{F}$ for constants $%
a $ and $b$, in accordance with $V_{\mu }$ being a gradient, as in (\ref%
{VectorAsGrad}) and (\ref{GradientCond}). \ This linear condition is
immediately integrated to obtain%
\begin{equation}
\mathcal{L}\left( B,F\right) =aB+L\left( F+bB\right) \ ,  \tag{C7}
\label{Lagrangian}
\end{equation}%
where $L\left( F+bB\right) $ is a differentiable function of the linear
combination $F+bB$. \ The field equations (\ref{FieldEqns}) are now%
\begin{equation}
\partial _{\mu }L^{\prime }=2\left( a+bL^{\prime }\right) V_{\mu }\ . 
\tag{C8}  \label{SimpleFieldEqns}
\end{equation}

As is well-known, there may be two distinct expressions for energy-momentum
tensors that result from any Lagrangian. \ From (\ref{Lagrangian}) the
canonical results for $\Theta _{\mu \nu }$, and its trace $\Theta =\Theta
_{\mu }^{\ \mu }$, are immediately seen to be%
\begin{eqnarray}
\Theta _{\mu \nu }^{\left[ \text{canonical}\right] } &=&\left( \partial
_{\mu }V_{\nu }\right) L^{\prime }-g_{\mu \nu }\left( aB+L\right) \ , 
\TCItag{C9}  \label{TCanon} \\
\Theta ^{\left[ \text{canonical}\right] } &=&FL^{\prime }-N\left(
aB+L\right) \ .  \notag
\end{eqnarray}%
Although not manifestly symmetric, it is nonetheless true that $\Theta _{\mu
\nu }^{\left[ \text{canonical}\right] }=\Theta _{\nu \mu }^{\left[ \text{%
canonical}\right] }$ \emph{on-shell} in light of the condition (\ref%
{GradientCond}). \ 

Surprisingly different results follow from covariantizing (\ref{Lagrangian})
with respect to an arbitrary background metric $g_{\mu \nu }$, varying the
action for $\sqrt{\left\vert \det g_{\alpha \beta }\right\vert }~\mathcal{L}$
with respect to that metric, and then taking the flat-space limit. \ This
procedure gives the \textquotedblleft gravitational\textquotedblright\
energy-momentum tensor and its trace:%
\begin{eqnarray}
\Theta _{\mu \nu }^{\left[ \text{gravitational}\right] } &=&-2\left(
a+bL^{\prime }\right) V_{\mu }V_{\nu }-g_{\mu \nu }\left( L-aB-\left(
F+2bB\right) L^{\prime }\right) \ ,  \TCItag{C10}  \label{TGrav} \\
\Theta ^{\left[ \text{gravitational}\right] } &=&\left( NF+\left(
2N-2\right) bB\right) L^{\prime }+\left( N-2\right) aB-NL\ .  \notag
\end{eqnarray}%
The unusual structure exhibited in this tensor follows because, as defined
by (\ref{VandF}), $V^{\mu }$ is a \href{https://en.wikipedia.org/wiki/Tensor_density}%
{relative contravariant vector} of weight $+1$ with no dependence on the
metric, so $\partial _{\mu }V^{\mu }$ is a relative scalar of weight $+1$,
also with no dependence on $g_{\mu \nu }$, and $V_{\mu }V^{\mu }=g_{\mu \nu
}V^{\mu }V^{\nu }$ is a relative scalar of weight $+2$ where all dependence
on the metric is shown explicitly. \ Hence the absolute scalar version of $%
\mathcal{L}\left( B,F\right) $ is given by 
\begin{equation}
\mathcal{L}=\frac{ag_{\mu \nu }V^{\mu }V^{\nu }}{\left\vert \det g_{\alpha
\beta }\right\vert }+L\left( \frac{\partial _{\mu }V^{\mu }}{\sqrt{%
\left\vert \det g_{\alpha \beta }\right\vert }}+\frac{bg_{\mu \nu }V^{\mu
}V^{\nu }}{\left\vert \det g_{\alpha \beta }\right\vert }\right) \ , 
\tag{C11}  \label{AbsoluteScalar}
\end{equation}%
where once again all the metric dependence is shown explicitly.

It is straightforward to check on-shell conservation of either (\ref{TCanon}%
) or (\ref{TGrav}), separately. \ However, it turns out the flat-space
equations of motion can now be written in the form (\ref{Elegant}) provided
a linear combination of $\Theta _{\mu \nu }^{\left[ \text{canonical}\right]
} $ and $\Theta _{\mu \nu }^{\left[ \text{gravitational}\right] }$ is used
for the system's energy-momentum tensor. \ Let%
\begin{equation}
\Theta _{\mu \nu }=\frac{N-2}{N-1}~\Theta _{\mu \nu }^{\left[ \text{canonical%
}\right] }+\frac{1}{N-1}~\Theta _{\mu \nu }^{\left[ \text{gravitational}%
\right] }\ .  \tag{C12}  \label{NiceT}
\end{equation}%
The trace is then%
\begin{equation}
\Theta =\Theta _{\mu }^{\ \mu }=2\left( F+bB\right) L^{\prime }-NL-\left(
N-2\right) aB\ .  \tag{C13}  \label{NiceTrace}
\end{equation}

The field equations (\ref{GradientCond}) and (\ref{SimpleFieldEqns}) give
for the left-hand side of (\ref{Elegant})%
\begin{equation}
\left( \square +m^{2}\right) V_{\mu }=\left( 1+\frac{m^{2}}{2}\frac{%
L^{\prime \prime }}{a+bL^{\prime }}\right) \partial _{\mu }\left(
F+bB\right) -b~\partial _{\mu }B\ ,  \tag{C14}
\end{equation}%
where (\ref{GradientCond}) implies $\square V_{\mu }=\partial ^{\lambda
}\partial _{\lambda }V_{\mu }=\partial ^{\lambda }\partial _{\mu }V_{\lambda
}=\partial _{\mu }F$. \ On the other hand, from (\ref{NiceTrace}) for any
constant $c$,%
\begin{equation}
c~\partial _{\mu }\Theta =c\left( _{\ }2\left( F+bB\right) L^{\prime \prime
}-\left( N-2\right) L_{\ }^{\prime }\right) \partial _{\mu }\left(
F+bB\right) -\left( N-2\right) ac~\partial _{\mu }B\ .  \tag{C15}
\end{equation}%
The choice $\left( N-2\right) ac=b$ reconciles the spurious $\partial _{\mu
}B$ term to give the desired form%
\begin{equation}
\left( \square +m^{2}\right) V_{\mu }=c~\partial _{\mu }\Theta  \tag{C16}
\label{ScaledFieldEqns}
\end{equation}%
\emph{provided} the function $L$ satisfies the second-order nonlinear
equation 
\begin{equation}
1+\frac{m^{2}}{2}\frac{L^{\prime \prime }\left( z\right) }{a+bL^{\prime
}\left( z\right) }=c\left( 2zL^{\prime \prime }\left( z\right) -\left(
N-2\right) L^{\prime }\left( z\right) \right) \ .  \tag{C17}  \label{ODE}
\end{equation}%
Moreover, the constant $c$ can be set to a convenient nonzero value by a few
scale changes. \ 

For example, if $\left( a,L\right) \ \rightarrow \ \left( \frac{ab}{2c},%
\frac{am^{2}}{2bc}L\right) $, along with the previous choice $\left(
N-2\right) ac=b\ \rightarrow \ a=2/\left( N-2\right) $, the equation for $L$
becomes%
\begin{equation}
1+\frac{m^{4}}{2b}\frac{L^{\prime \prime }}{b+m^{2}L^{\prime }}=\frac{m^{2}}{%
b}\left( \frac{2}{\left( N-2\right) }~zL^{\prime \prime }\left( z\right)
-L^{\prime }\left( z\right) \right) \ .  \tag{C18}
\end{equation}%
The rescaling $z\rightarrow m^{2}w/b$ then gives%
\begin{equation}
1+\frac{1}{2}\frac{L^{\prime \prime }}{1+L^{\prime }}=\frac{2}{\left(
N-2\right) }~wL^{\prime \prime }-L^{\prime }\ ,  \tag{C19}  \label{FinalODE}
\end{equation}%
where the $^{\prime }$s in (\ref{FinalODE}) are $\frac{d}{dw}$s. \ The
solutions of this final differential equation are very dependent upon $N$. \
With the initial condition $L^{\prime }\left( 0\right) =0$ a first integral
is given by 
\begin{equation}
L^{\prime }=X-1\ ,  \tag{C20}
\end{equation}%
where $X$ is a root of 
\begin{equation}
X^{\frac{N}{N-2}}=\left( 1-\frac{2N}{\left( N-2\right) }wX\right) \ , 
\tag{C21}  \label{GetRoots}
\end{equation}%
such that $X\rightarrow 1$ as $w\rightarrow 0$. \ 

The simplest\ cases of (\ref{GetRoots}) are for $N=4$ and $N\rightarrow
\infty $. \ In those cases, $\left. \left( X\right) ^{\frac{N}{N-2}}-\left(
1-\frac{2N}{\left( N-2\right) }wX\right) \right\vert _{N=4}=X^{2}-1+4wX$, so
for $N=4$ the roots are: $\left\{ X=\frac{1}{2}\left( -4w+2\sqrt{4w^{2}+1}%
\right) \right\} $ and $\left\{ X=\frac{1}{2}\left( -4w-2\sqrt{4w^{2}+1}%
\right) \right\} $; while $N\rightarrow \infty $ gives just $X=\left(
1-2wX\right) $, whose solution is: $\left\{ X=\frac{1}{1+2w}\right\} $. \
Thus%
\begin{eqnarray}
\left. L^{\prime }\left( w\right) \right\vert _{N=4} &=&\frac{1}{2}\left( 2%
\sqrt{1+4w^{2}}-2-4w\right)  \TCItag{C22}  \label{L'Four} \\
\left. L^{\prime }\left( w\right) \right\vert _{N\rightarrow \infty } &=&%
\frac{-2w}{1+2w}  \TCItag{C23}  \label{L'Infinity}
\end{eqnarray}%
For these two special cases a final integration with the initial condition $%
L\left( 0\right) =0$ gives%
\begin{eqnarray}
\left. L\left( w\right) \right\vert _{N=4} &=&-w-w^{2}+\frac{1}{2}w\sqrt{%
1+4w^{2}}+\frac{1}{4}\ln \left( 2w+\sqrt{1+4w^{2}}\right) \ ,  \TCItag{C24}
\label{LFour} \\
\left. L\left( w\right) \right\vert _{N\rightarrow \infty } &=&-w+\frac{1}{2}%
\ln \left( 1+2w\right) \ .  \TCItag{C25}  \label{LInfinity}
\end{eqnarray}%
The first of these reproduces the result in \cite{C2019}.

For other $N$ it \emph{might} seem that things can get out of hand, except
perhaps for $N=3$, $6$, and $8$. \ For the first two of these cases, (\ref%
{GetRoots}) results in a cubic equation, which is tractable. \ For $N=8$, (%
\ref{GetRoots}) is a quartic equation, which is also tractable. \ But for
other $N$, (\ref{GetRoots}) is quintic, or worse. \ 

The general solution for these other values of $N$ is indeed nontrivial, but
the Taylor series for $L^{\prime }$ is remarkably simple. \ For example, 
\begin{eqnarray}
L^{\prime }\left( w\right) &=&-2w+4\frac{\left( N-3\right) w^{2}}{\left(
N-2\right) }-\frac{8}{3}\left( N-4\right) \left( 3N-8\right) \frac{w^{3}}{%
\left( N-2\right) ^{2}}  \TCItag{C26} \\
&&+\frac{8}{3}\left( N-5\right) \left( 2N-5\right) \left( 3N-10\right) \frac{%
w^{4}}{\left( N-2\right) ^{3}}  \notag \\
&&-\frac{16}{15}\left( N-6\right) \left( 2N-6\right) \left( 3N-12\right)
\left( 5N-12\right) \frac{w^{5}}{\left( N-2\right) ^{4}}+O\left(
w^{6}\right) \ .  \notag
\end{eqnarray}%
As a polynomial in $N$, the coefficient of $w^{m+1}/\left( N-2\right) ^{m}$ 
\emph{always} factors over the rationals. \ The complete series is \cite%
{TSvK}%
\begin{equation}
L^{\prime }\left( w\right) =\frac{N-2}{N}\sum_{m=1}^{\infty }\frac{1}{m!}%
\frac{\Gamma \left( \frac{N-2}{N}\left( 1+m\right) \right) }{\Gamma \left( 2-%
\frac{2}{N}\left( 1+m\right) \right) }\left( \frac{2Nw}{2-N}\right) ^{m}\ . 
\tag{C27}
\end{equation}%
This result for $L^{\prime }$ is a special case of Fox's generalized
confluent hypergeometric function \cite{Fox}. \ A final integration then
yields the sought-for $\mathcal{L}$ for any $N$.

The field strength for the dual scalar obeys an equation that can be
manipulated in a manner similar to that used in the main text to relate the
dual gravitational field to the ND\ Ogievetsky-Polubarinov model. \ Taking
the divergence of (\ref{Elegant}) gives%
\begin{equation}
\left( \square +m^{2}\right) F=\kappa ~\square \Theta \ .  \tag{C28}
\label{DualFN}
\end{equation}%
A nonlinear field redefinition, namely,%
\begin{equation}
\Psi =\frac{1}{m^{2}}\left( F-\kappa \Theta \right) \ ,  \tag{C29}
\label{NDFNReDefn}
\end{equation}%
then converts (\ref{DualFN}) into%
\begin{equation}
\left( \square +m^{2}\right) \Psi =-\kappa \Theta \ .  \tag{C30}
\label{NDFN}
\end{equation}%
This is the field equation for the ND extended Freund-Nambu model \cite{FN}
of a fundamental scalar field $\Psi $ coupled to the trace of its own
energy-momentum tensor. \ That is to say, the $V_{\alpha _{1}\cdots \alpha
_{N-1}}$ model constructed here is the massive dual of the Freund-Nambu
scalar theory on-shell. \ In the latter model, of course, the trace\ is
expressed as a local functional of $\Psi $, whereas $\Theta $ in (\ref{NDFN}%
) is a functional of $V_{\alpha _{1}\cdots \alpha _{N-1}}$ that must be
re-expressed in terms of $\Psi $. \ That this can be done is perhaps not
obvious, but nonetheless it is true.

Given the structural similarities between the Freund-Nambu theory and scalar
gravitation \cite{Deser}, it is perhaps more plausible that the complete
Lagrangian\ for the self-coupled $T_{\left[ \lambda _{1}\cdots \lambda _{N-2}%
\right] \mu }$\ field can be determined to all orders in $\kappa $.


\newpage

\begin{thebibliography}{99}
\bibitem{FMS1968} P.G.O. Freund, A. Maheshwari, and E. Schonberg,
\textquotedblleft Finite Range Gravitation\textquotedblright\ \href{http://adsabs.harvard.edu/doi/10.1086/150118}%
{Astro.Journal 157 (1969) 857--867}.

\bibitem{P} J.B. Pitts and W.C. Schieve, \textquotedblleft Universally
coupled massive gravity\textquotedblright\ \href{https://doi.org/10.1007/s11232-007-0055-7}%
{Theor.Math.Phys. 151 (2007) 700-717}; J.B. Pitts, \textquotedblleft
Universally Coupled Massive Gravity, II: Densitized Tetrad and Cotetrad
Theories\textquotedblright\ \href{https://doi.org/10.1007/s10714-011-1280-9}{%
Gen.Rel.Grav. 44 (2012) 401-426}; J.B. Pitts, \textquotedblleft Universally
coupled massive gravity, III: dRGT--Maheshwari pure spin-2,
Ogievetsky--Polubarinov and arbitrary mass terms\textquotedblright\ \href{https://doi.org/10.1016/j.aop.2015.12.002}%
{Ann.Phys. 365 (2016) 73-90}.

\bibitem{C1980} T.L. Curtright, \textquotedblleft Generalized Gauge
Fields\textquotedblright\ \href{https://doi.org/10.1016/0370-2693(85)91235-3}%
{Phys.Lett. 165B (1985) 304-308}; T.L. Curtright and P.G.O. Freund,
\textquotedblleft Massive Dual Fields\textquotedblright\ \href{https://doi.org/10.1016/0550-3213(80)90174-1}%
{Nucl.Phys. B172 (1980) 413-424}.

\bibitem{C2019} T.L. Curtright, \textquotedblleft Massive Dual Spinless
Fields Revisited\textquotedblright\ \href{https://arxiv.org/abs/1907.11530}{%
arXiv:1907.11530 [hep-th]}; T.L. Curtright and H. Alshal, \textquotedblleft
Massive Dual Spin 2 Revisited\textquotedblright\ \href{https://arxiv.org/abs/1907.11532}%
{arXiv:1907.11532 [hep-th]}.

\bibitem{C1986} T.L. Curtright and C.B. Thorn, \textquotedblleft Symmetry
Patterns in the Mass Spectra of Dual String Models\textquotedblright\ \href{https://doi.org/10.1016/0550-3213(86)90525-0}%
{Nucl.Phys. B274 (1986) 520-558}; T.L. Curtright, C.B. Thorn, and J.
Goldstone, \textquotedblleft Spin Content of the Bosonic
String\textquotedblright\ \href{https://doi.org/10.1016/0370-2693(86)90329-1}%
{Phys.Lett. B175 (1986) 47-52}; T.L. Curtright, G.I. Ghandour, and C.B.
Thorn, \textquotedblleft Spin Content of String Models\textquotedblright\ 
\href{https://doi.org/10.1016/0370-2693(86)91076-2}{Phys.Lett. B182 (1986)
45-52}; T.L. Curtright, \textquotedblleft Counting Symmetry Patterns in the
Spectra of Strings\textquotedblright\ \href{http://inspirehep.net/record/234762?ln=en}%
{SUNY STONY BROOK - ITP-SB-86-74}, pp 304-333 in \textit{String Theory,
Quantum Cosmology and Quantum Gravity, Integrable and Conformal Invariant
Theories}, Proceedings of the Paris-Meudon Colloquium, 22-26 September 1986,
H. J. De Vega (Author), N. Sanchez (Author, Editor), World Scientific 1987.
ISBN-13: 978-9971502867.

\bibitem{H} C.M. Hull, \textquotedblleft Strongly coupled gravity and
duality\textquotedblright , \href{https://doi.org/10.1016/S0550-3213(00)00323-0}%
{Nucl.Phys. B583 (2000) 237-259}; C.M. Hull, \textquotedblleft Duality in
gravity and higher spin gauge fields\textquotedblright\ \href{https://arxiv.org/abs/hep-th/0107149}%
{JHEP 0109 (2001) 027}.

\bibitem{W} P.C. West, \textquotedblleft $E_{11}$ and $M$-theory%
\textquotedblright\ \href{https://iopscience.iop.org/article/10.1088/0264-9381/18/21/305}%
{Class.Quant.Grav. 18 (2001) 4443}; \ P. West, \textquotedblleft A brief
review of E theory\textquotedblright\ \href{http://dx.doi.org/10.1142/S0217751X1630043X}%
{Int.J.Mod.Phys. A 31 (2016) 1630043}.

\bibitem{DHN} T. Damour, M. Henneaux, H. Nicolai, \textquotedblleft $E_{10}$
and a Small Tension Expansion of M theory\textquotedblright\ \href{https://doi.org/10.1103/PhysRevLett.89.221601}%
{Phys.Rev.Lett. 89 (2002) 221601}.

\bibitem{D} A. Danehkar, \textquotedblleft Electric-Magnetic Duality in
Gravity and Higher-Spin Fields\textquotedblright\ \href{https://doi.org/10.3389/fphy.2018.00146}%
{Front.Phys. 09 January 2019}.

\bibitem{GKMU} B. Gonz\'{a}lez, A. Khoudeir, R. Montemayor, and L.F.
Urrutia, \textquotedblleft Duality for massive spin two theories in
arbitrary dimensions\textquotedblright\ \href{https://doi.org/10.1088%2F1126-6708%2F2008%2F09%2F058}%
{JHEP 0809 (2008) 058}.

\bibitem{OP} V.I. Ogievetsky and I.V. Polubarinov, \textquotedblleft
Interacting field of spin 2 and the Einstein equations\textquotedblright\ 
\href{https://doi.org/10.1016/0003-4916(65)90077-1}{Ann.Phys. 35 (1965)
167--208}.

\bibitem{Hamermashed} D. Hilbert, \textquotedblleft \"{U}ber die Theorie der
algebraischen Formen\textquotedblright\ \href{https://link.springer.com/article/10.1007/BF01208503}%
{Math.Ann. 36 (1890) 473-534}. \newline
J.J. Sylvester, \textquotedblleft On a Theory of Syzygetic Relations
...\textquotedblright\ \href{https://doi.org/10.1098/rstl.1853.0018}{%
Philos.Trans.Roy.Soc.London 143\ (1853) 407-548}.

\bibitem{TSvK} T.S. Van Kortryk, \href{https://cgc.physics.miami.edu/GetRoots.pdf}%
{unpublished}.

\bibitem{Fox} C. Fox, \textquotedblleft The asymptotic expansion of integral
functions defined by generalized hypergeometric series\textquotedblright\ 
\href{https://doi.org/10.1112/plms/s2-27.1.389}{Proc.London Math.Soc. 27
(1928) 389-400}.

\bibitem{FN} P.G.O. Freund and Y. Nambu \textquotedblleft Scalar Fields
Coupled to the Trace of the Energy-Momentum Tensor\textquotedblright\ \href{https://doi.org/10.1103/PhysRev.174.1741}%
{Phys.Rev. 174 (1968) 1741-1743}.

\bibitem{Deser} S. Deser and L. Halpern, \textquotedblleft Self-coupled
scalar gravitation\textquotedblright\ \href{http://dx.doi.org/10.1007/BF00756892}%
{Gen.Rel.Grav. 1 (1970) 131-136}.
\end{thebibliography}
\end{document}